\documentclass[
    amsmath,
    amssymb,
    reprint,
    aps,
    prx,
    superscriptaddress,
    longbibliography
]{revtex4-1}

\usepackage[utf8]{inputenc}
\usepackage[T1]{fontenc}
\usepackage{physics}
\usepackage{nicefrac}
\usepackage{bbm, bm}
\usepackage{makecell}
\usepackage{multirow}
\usepackage{booktabs}
\usepackage{overpic}
\usepackage{tikz}
\usepackage{pgfplots}
\pgfplotsset{compat=1.18}
\usepackage[dvipsnames]{xcolor}

\usepackage[breaklinks=true, colorlinks]{hyperref}
\hypersetup{linkcolor=blue, citecolor=blue, urlcolor=MidnightBlue}
\usepackage{cleveref}

\begin{document}

\title{Adiabatic dressing of quantum enhanced Markov chains}

\author{Wen Ting Hsieh}
\thanks{These authors contributed equally to this work}
\affiliation{Center for Quantum Phenomena, Department of Physics, New York University, 726 Broadway, New York, New York 10003, USA}

\author{Alev Orfi}
\thanks{These authors contributed equally to this work}
\affiliation{Center for Quantum Phenomena, Department of Physics, New York University, 726 Broadway, New York, New York 10003, USA}
\affiliation{Center for Computational Quantum Physics, Flatiron Institute, 162 5th Avenue, New York, NY 10010, USA}

\author{Dries Sels}
\affiliation{Center for Computational Quantum Physics, Flatiron Institute, 162 5th Avenue, New York, NY 10010, USA}
\affiliation{Department of Physics, Boston University, 590 Commonwealth Ave., Boston, Massachusetts 02215, USA}

\date{\today}

\begin{abstract}
Quantum-enhanced Markov chain Monte Carlo, a hybrid quantum-classical algorithm in which configurations are proposed by a quantum proposer and accepted or rejected by a classical algorithm, has been introduced as a possible method for robust quantum speedup. Previous work has identified competing factors that limit the algorithm's performance: the quantum dynamics should delocalize the system across a range of classical states to propose configurations beyond the reach of simple classical updates, whereas excessive delocalization produces configurations unlikely to be accepted, slowing the chain's convergence. Here, we show that controlling the degree of delocalization by adiabatically dressing the quench protocol can significantly enhance the Markov gap in paradigmatic spin-glass models. 
    
\end{abstract}

\maketitle

\section{Introduction}
Although current quantum processors continue to grow in size and gate capacity, their practical utility remains constrained by noise and limited coherence time. These limitations have motivated the search for algorithms that are both applicable to practical problems and retain performance under imperfect operations. One relevant task is sampling from Boltzmann distributions of classical Hamiltonians, which appears throughout statistical mechanics, combinatorial optimization, and probabilistic machine learning. 

Consider a system of $N$ binary variables $x_i = \pm 1$, described by a classical Hamiltonian $H_c$. At inverse temperature $\beta$, configurations are distributed according to the Boltzmann distribution,
\begin{equation}
    \pi(x) = \frac{1}{\mathcal{Z}}e^{-\beta H_c(x)},
\end{equation}
where the partition function $ \mathcal{Z}$, is typically computationally intractable. As a result, sampling from $\pi$ is often the only practical method to estimate the thermodynamic properties of the system. 

Quantum-enhanced Markov chain Monte Carlo is a near-term approach for Boltzmann sampling that is robust to noise as it preserves convergence to the target distribution even when the underlying quantum subroutine is imperfect \cite{layden2023quantum}. This hybrid algorithm constructs a Markov chain on the classical configuration space, in which candidate updates are proposed by measuring the outcome of a quantum evolution. A classical step then accepts or rejects the proposal, ensuring convergence to the target stationary distribution. 

Variants of this approach have been explored, with different quantum proposal strategies and classical processing \cite{scriva2023accelerating,lockwood2024quantum,nakano2024markov,arai2025quantum,christmann2025quantum,ferguson2025quantum, nakano2025fair,nakano2026neural,ferguson2026methods,marshall2026quantum, Metcalf2022Quantum}. Ideal performance occurs when the quantum subroutine proposes configurations that are close in energy to the previous state of the chain, therefore leading to efficient chain mixing, but sufficiently far in Hamming distance to outperform local classical strategies \cite{orfi2024barriers}. This algorithm has been deployed as a sampling subroutine \cite{wilson2024non} and applied to problems such as causal-set sampling \cite{ferguson2025dynamics}, cosmological inference \cite{sarracino2025quantum}, and variational Monte Carlo \cite{chang2025quantum}. Building on prior analyses of the performance of this class of algorithms \cite{orfi2024bounding,orfi2024barriers,christmann2025quantum}, we introduce an adiabatically dressed quantum evolution that substantially improves performance as compared to the original quench protocol.

\section{Methods}
Markov chain Monte Carlo (MCMC) provides a general framework for sampling by constructing a stochastic process that equilibrates to a target distribution. A common implementation of such a Markov chain consists of two stages. First, a candidate configuration $x$ is proposed with probability $Q(x|y)$ conditioned on the current state $y$. Second, the proposed configuration is accepted as the new state of the chain with probability $A(x|y)$, which depends on the target distribution. The process results in a stochastic transition matrix $P$, defined as 
\begin{equation*}
    P(y,x) = \begin{cases}
     Q(x|y)A(x|y), & \text{if}\ x\neq y\\
     1-\sum_{z\neq y} Q(z|y)A(z|y), & \text{if}\ x=y
    \end{cases}
\end{equation*}
whose elements $P(y,x)$ denote the probability of transitioning from configuration $y$ to $x$. A common choice, and what will be considered here, is the Metropolis-Hastings acceptance probability,
\begin{equation}\label{eq:MH_accept}
        A(x|y)  = \min\left(1,e^{-\beta(H_c(x)-H_c(y))}\frac{Q(y|x)}{Q(x|y)}\right).
\end{equation}
An important performance metric for any MCMC method is the mixing time $t_{\mathrm{mix}}$, the number of steps required for the chain's distribution to be close to the stationary distribution in total variation distance. As the dynamics is generated through the repeated application of the transition matrix, convergence is determined by the spectral properties of $P$. The stationary distribution $\pi$ corresponds to the unit eigenvalue of $P$, while the convergence is determined by the spectral gap, $\delta = 1 -|\lambda_2|$, where $\lambda_2$ is the second largest eigenvalue. The mixing time is related to this spectral gap through the bound,
\begin{equation}
   (\delta^{-1}-1)\ln\left(\frac{1}{2\epsilon}\right)\leq t_{\text{mix}} \leq \delta^{-1}\ln\left(\frac{1}{\epsilon \pi_{\min}}\right)
\end{equation}
where $\pi_{\min} = \min_{x\in S}\pi(x)$ \cite{levin_MarkovChainsMixingTime}. 

While the spectral gap is a valuable performance metric, computing $\delta$ requires diagonalizing an exponentially large transition matrix, restricting such analyses to small system sizes. Analyzing mixing bottlenecks provides an alternative route to characterizing the performance of the chain \cite{levin_MarkovChainsMixingTime}, an approach that has previously provided insight into the properties of quantum-enhanced MCMC \cite{orfi2024bounding,orfi2024barriers}. Specifically, the equilibrium flow between a subset of configurations $S$ and its complement is,
\begin{equation}
    E(S,S^c) = \sum_{x\in S, y\in S^c}\pi(x)P(x,y).
\end{equation}
The spectral gap can be upper bounded by minimizing the equilibrium flow across possible subsets \cite{levin_MarkovChainsMixingTime}, 
\begin{equation}\label{eq:bottleneck_UB}
    \delta \leq \min_{S: \pi(S)\leq 1/2} \frac{E(S,S^c)}{\pi(S)\pi(S^c)}.
\end{equation}
While the minimization is generally intractable, it can be relaxed by evaluating the expression on any fixed subset $B$, providing the potentially looser bound,
\begin{equation}\label{eq:lambda}
    \delta \leq \Lambda(B)
    \equiv \frac{E(B,B^c)}{\pi(B)\pi(B^c)}.
\end{equation}
As the focus is on sampling the low-temperature Boltzmann distribution, choosing configurations in $B$ with higher energy than those in $B^c$ can yield a tight bound and provide insight into the mixing. 

In the original quantum-enhanced MCMC formulation, classical configurations are evolved under a Hamiltonian comprised of the classical Hamiltonian and a transverse field of strength $h$, and then measured in the computational basis \cite{layden2023quantum}. This procedure leads to a proposal probability,
\begin{equation}
    Q(x|y) = |\bra{x}U\ket{y}|^2,
\end{equation}
where $U = e^{-iHt}$. In Ref.~\cite{orfi2024barriers}, a bottleneck analysis linked the mixing time of the algorithm to how strongly the eigenstates of $H$ are localized in the classical configuration basis. In particular, if these eigenstates are fully delocalized, the quantum algorithm has the same spectral gap scaling as a uniform classical proposal strategy. It was found that the best performance occurred at a problem-specific optimal field $h_{\mathrm{opt}}$, where $h$ was large enough such that proposals are sufficiently non-local. At larger transverse fields, the eigenstates become increasingly delocalized, and the chain’s performance degrades correspondingly. Here, we propose to control the extent of eigenstate delocalization by replacing a quench with a slow ramp of the transverse field, thereby mitigating the known performance limitations.

A time-dependent variant of quantum-enhanced MCMC was also considered in Ref.~\cite{arai2025quantum}, where an annealing schedule biases proposals toward configurations near the classical ground state. However, in this annealing case, the proposal probability is no longer symmetric. The acceptance probability of Eq.~\ref{eq:MH_accept} requires the ratio of the proposal probabilities, which cancel if the proposal is symmetric. In the absence of this symmetry, this ratio must be calculated at each step to ensure convergence, which is intractable if a quantum proposal is used. 

\begin{figure}
    \centering
    \hspace*{-0.3cm} 
    \begin{tikzpicture}
\begin{axis}[
    width=8.5cm,height=3.5cm,
    axis lines=left,
    xmin=0,xmax=6.2,
    ymin=0,ymax=1.25,
    xlabel={$t$},
    ylabel={$\gamma(t)$},
    xlabel style={
        at={(axis description cs:1,0)},
        anchor=west
    },
    xtick=\empty,
    ytick={0,1},
    yticklabels={$0$,$1$},
    clip=false,
]
\def\a{1.0}  
\def\k{4.0}  
\def\T{6.0} 
\addplot[thick,blue,domain=0:\a,samples=200]
({x},{(sin(deg(pi/2*sin(deg(pi*x/(2*\a)))^2)) )^2});
\addplot[thick,blue,domain=\a:\a+\k] ({x},{1});
\addplot[thick,blue,domain=\a+\k:\T,samples=200]
({x},{(sin(deg(pi/2*sin(deg(pi*(\T-x)/(2*\a)))^2)) )^2});
\addplot[dashed] coordinates {(\a,0) (\a,1)};
\addplot[dashed] coordinates {(\a+\k,0) (\a+\k,1)};
\draw[<->] (axis cs:0,-0.18) -- (axis cs:\a,-0.18)
    node[midway,below] {$\alpha$};
\draw[<->] (axis cs:\a,-0.18) -- (axis cs:\a+\k,-0.18)
    node[midway,below] {$\kappa$};
\draw[<->] (axis cs:\a+\k,-0.18) -- (axis cs:\T,-0.18)
    node[midway,below] {$\alpha$};
\end{axis}
\end{tikzpicture}
    \caption{Three-stage ramp protocol for the Hamiltonian parameter $\gamma(t)$. A smooth ramp-up from 0 to 1 over a duration $\alpha$ is defined by Eq.~\ref{eq:ramp}, followed by a plateau of length $\kappa$, and a ramp-down obtained by time-reversing the ramp-up.}
    \label{fig:rampDiagram}
\end{figure}
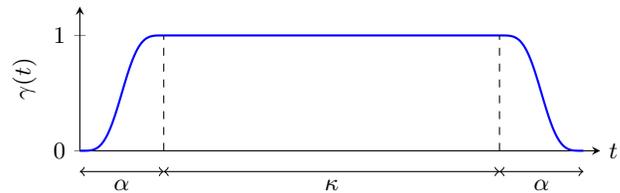

Instead, we consider a time-dependent Hamiltonian of the form,
\begin{equation}
    H(t) = H_c + \gamma(t) h\sum_{i=1}^N \sigma^x_i,
    \label{eq:mixingH}
\end{equation}
where $\gamma(t)$ follows a three-stage schedule, as illustrated in Fig.~\ref{fig:rampDiagram}. This time-symmetric protocol yields a symmetric proposal probability such that the Metropolis-Hastings acceptance step remains efficiently computable. This approach improves performance while preserving the essential feature of the original quantum-enhanced MCMC scheme, convergence to the target distribution even when the quantum proposal is computed imperfectly.

\section{Performance}
The effectiveness of any MCMC proposal strategy depends on the problem considered. For example, in the marked-item sampling problem, no unital quantum proposal can have any improvement over a uniform classical strategy \cite{orfi2024bounding}. For that problem, this ramped method has identical performance to the quenched proposal. Here, we instead consider more structured problems, for which a speedup is not ruled out, and characterize the resulting mixing behaviour through a combination of bottleneck analysis and numerical investigation.

Different ramp protocols yield comparable performance, as detailed in Appendix~\ref{appendix:ramps}. However, smoother schedules reduce diabatic excitations and therefore diminish finite-size artifacts in small systems~\cite{sels2017minimizing}. For the three-stage schedule in Fig.~\ref{fig:rampDiagram}, we define a smooth ramp-up function $\gamma_1(t)$ on $t\in[0,\alpha]$ by, 
\begin{equation}\label{eq:ramp}
    \gamma_1(t) = \sin^2\left(\frac{\pi}{2}\sin^2\left(\frac{\pi t}{2\alpha}\right)\right).
\end{equation}
The full protocol additionally contains a plateau of length $\kappa$ during which $\gamma(t)=1$, followed by a ramp-down equal to the time-reversal of $\gamma_1$. 

To isolate the effect of introducing a ramp relative to the quench proposal, we focus on the mixing-time scaling in the large-$\kappa$ limit, assuming the system equilibrates during the plateau. This regime can be directly compared to the long-time limit of the quench protocol \cite{orfi2024barriers}. A detailed analysis of the $\kappa$ dependence for the disordered models is presented together with the scaling behavior of the spectral gap. In the large-$\kappa$ limit, the proposal probability equilibrates to the time-averaged probability,
\begin{equation}
    Q(x|y) = \sum_n |\langle n|U_1|x\rangle|^2|\langle n|U_1|y\rangle|^2,
\end{equation}
where $\{|n\rangle\}$ are eigenstates of the plateau Hamiltonian with $\gamma=1$, and $U_1$ is the unitary evolution during the ramp segment $\gamma_1(t)$. The spectral gap of this time-averaged protocol upper bounds the $\kappa$-averaged spectral gap of the $\kappa$-dependent protocol \cite{orfi2024barriers}, and the numerical studies of the spin-glass system presented below show that the two exhibit similar scaling. 

For a subset $B$ whose configurations have larger energy than those in $B^c$, the bottleneck bound of Eq.~\ref{eq:lambda} has the following form,
\begin{equation}
    \Lambda(B) = \frac{1}{\pi(B)\pi(B^c)}\sum_{\substack{n,\; x\in B\\ y\in B^c}} \pi(x) |\langle n |U_1|x\rangle|^2 |\langle n |U_1|y\rangle|^2.
\end{equation}
The bound highlights a key advantage over the quench case: the ramp protocol allows control over eigenstate overlaps and thereby can circumvent the unfavourable scaling induced by excessive eigenstate delocalization. More precisely, evolution under $U_1$ spreads a classical configuration over an energy window in the $\{|n\rangle\}$ basis, where the width of the window depends on the ramp duration $\alpha$. Starting from the quench limit at $\alpha =0$, where $U_1=1$, the spectral gap increases with $\alpha$ (at sufficiently large $h$), corresponding to a localization of the proposed configurations in energy. At the other extreme, when $\alpha \rightarrow \infty$, the evolution becomes adiabatic, and each classical configuration $|x\rangle$ gets transported into a particular quantum state $|n\rangle$. Due to the time-reversal symmetry of the dynamics, the evolution returns to the original state, preventing the chain from mixing. The best performance occurs in an intermediate regime, where diabatic transitions allow the proposal of a broader set of states while keeping the energy window sufficiently narrow. 

This energy localization is particularly useful when $h$ exceeds the optimal quench value $h_{\mathrm{opt}}$. In this regime, the eigenstates of the quantum Hamiltonian differ substantially from those of $H_c$, enabling proposals that are far in Hamming distance. These types of proposals are often difficult for classical methods to generate. In the quench strategy, this eigenstate difference generates a broad proposal distribution over classical configurations, including high-energy proposals, leading to poor mixing. Introducing the control parameter $\alpha$ allows the ramp protocol to localize proposals in energy, without affecting the structure of the eigenstates. These qualitative mixing trends are supported by analytical results on the Ising chain and by numerical investigations of spin-glass systems, as shown in the next sections.

\subsection{Ising Chain}
\begin{figure}[t]
    \centering
    \includegraphics[width=\linewidth, trim=0.2cm 0.32cm 0 0.2cm, clip]{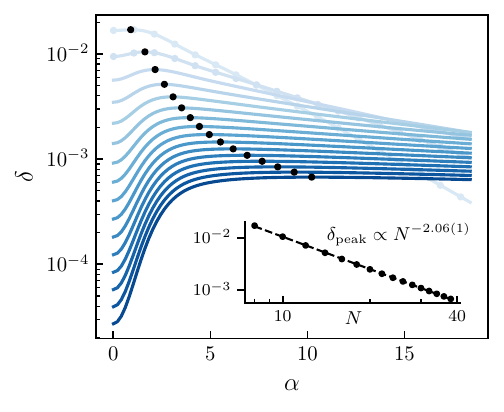}
    \caption{Bottleneck bound on the spectral gap $\delta$ for sampling the Ising chain at $\beta=5$ and fixed transverse field $h=1.5$. Solid lines show the bound for $N=8-40$, while dots denote exact gaps for smaller systems. The inset shows the system-size dependence of the peak gap, exhibiting polynomial scaling.}
    \label{fig:ising_sin_scaling}
\end{figure}

Consider the $N$-site Ising chain with periodic boundary conditions,
\begin{equation}\label{eq:ising}
    H_c = -\sum_{i=1}^N\sigma_i^z\sigma_{i+1}^z.
\end{equation}
Previous results have shown that the bottleneck bound of Eq.~\ref{eq:bottleneck_UB} is tight for this model at low temperatures if $B$ is chosen to be the first-excited state manifold  \cite{orfi2024barriers}. Details of the application of this bound to this time-dependent case are provided in Appendix~\ref{appendix:ising_UB}. This approach allows substantially larger system sizes to be analyzed than what would be accessible by computing $\delta$ through exact diagonalization. Fig.~\ref{fig:ising_sin_scaling} displays the bottleneck bound (solid lines) for $N=8-40$ as a function of the ramp time $\alpha$ at fixed transverse field ($h=1.5$) for low-temperature ($\beta=5$) sampling. Comparison with the exact spectral gap (dots) for smaller systems confirms the accuracy of the bound.

Increasing $\alpha$ from the quench limit ($\alpha = 0$) increases the spectral gap and alters the scaling with the system size. At sufficiently large $\alpha$ (evident here for small system sizes), the dynamics becomes increasingly adiabatic, increasing the probability to ultimately propose to stay in the original state and thereby decreasing the spectral gap. As the system size increases, the range of $\alpha$ that yields the best spectral gap broadens. 

In this example, the transverse field exceeds the optimal quench transverse field, $h_{\mathrm{opt}}$, corresponding to the regime in which adiabatic dressing can improve upon the quench protocol. By contrast, at weaker transverse fields, the quenched gap is already limited by poor mixing with other states, and adiabatic dressing is expected to exacerbate this effect. This is confirmed analytically in Appendix~\ref{appendix:ising_linear} for the case of a linear ramp, whose dynamics can be solved analytically.

The inset in Fig.~\ref{fig:ising_sin_scaling} shows the scaling of the peak gap, $\delta_{\mathrm{peak}}$, with system size, reproducing the optimal $\mathcal{O}(N^{-2})$ scaling of a classical local spin-flip MCMC method. Further details are given in Appendix~\ref{appendix:ising_linear} for the analytically solvable linear-ramp case, and in Appendix~\ref{appendix:ising_LZ}, where the same scaling is recovered from a more general Landau-Zener argument applicable to the ramp considered here. This result highlights a clear contrast with the quench protocol, which exhibits exponential scaling at any fixed $h$ for this model \cite{orfi2024barriers}. The ramp time provides an additional control parameter that regulates the quantum proposal and can qualitatively improve the scaling of the algorithm.

\subsection{Infinite-Range Spin Glasses}

The integrability of the Ising chain makes it possible to analyze the performance of this method at large system sizes, but its efficient mixing under local spin-flip updates makes it unrepresentative of harder problems. Spin glasses provide a more relevant setting, as their many metastable states lead to exponentially slow mixing at low temperatures \cite{barahona1982computational}. The Sherrington-Kirkpatrick (SK) model is therefore considered as a representative fully connected spin glass with frustrated couplings,
\begin{equation}
H_c = \sum_{i<j
}J_{ij}\sigma_i^z\sigma_{j}^z + \sum_ih_{i}\sigma_i^z, 
\end{equation}
where the couplings are drawn from  $J_{ij} \sim \mathcal{N}\left(0, 1/N\right) $ \cite{sherrington1975solvable}. In addition, the local fields $h_{i}$ are drawn from a uniform distribution $h_{i} \sim \mathcal{U}(-0.25, 0.25)$ to break the model's $\mathbb{Z}_2$ symmetry. To test the robustness of these results across distinct spin-glass phases, the 3-spin model is also investigated,
\begin{equation} 
H_c = \sum_{i<j<k
}J_{ijk}\sigma_i^z\sigma_{j}^z\sigma_{k}^z,
\end{equation}
with couplings drawn from $ J_{ijk} \sim \mathcal{N}\left(0,3/N^2\right)$ \cite{gardner1985spin}. Unlike the SK model, which exhibits a full replica-symmetry-breaking (RSB) phase at low-temperatures, the 3-spin model undergoes a first-order transition into a 1-RSB phase \cite{mezard1984nature,gardner1985spin}, which may lead to different chain dynamics.

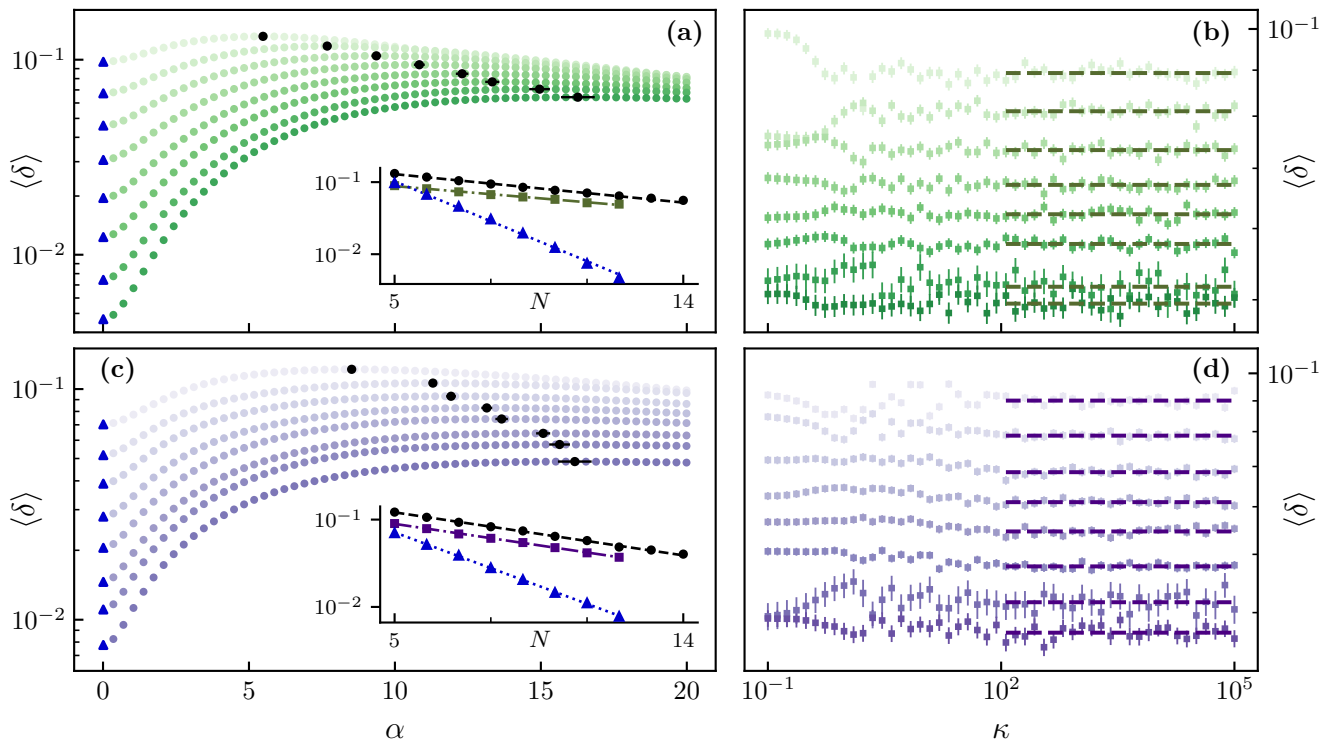
\begin{figure*}[t]
    \centering
    \begin{overpic}[width=\linewidth, trim=0 0.3cm 0 0, clip]{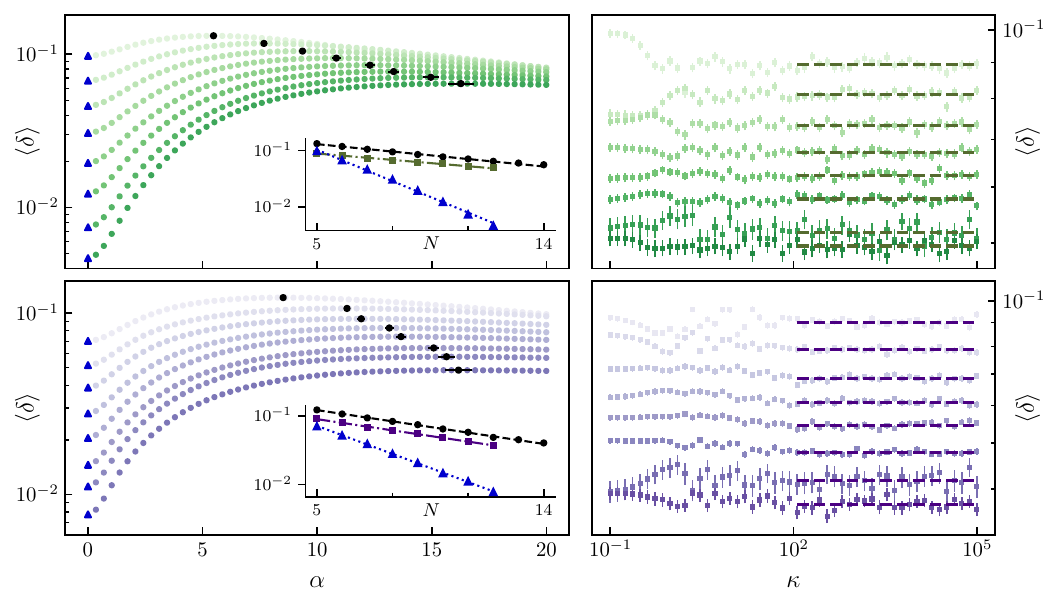}
        \put(50, 52){\fontsize{10}{12}\selectfont\textbf{(a)}}   
        \put(89, 52){\fontsize{10}{12}\selectfont\textbf{(b)}}
        \put(8, 27){\fontsize{10}{12}\selectfont\textbf{(c)}}
        \put(89, 27){\fontsize{10}{12}\selectfont\textbf{(d)}}
    \end{overpic}
    \caption{Spectral gap of the adiabatically dressed MCMC protocol at $\beta = 5$ and $h=1.5$ for the SK (green) and 3-spin (purple) models. Panels (a) and (c) show the disordered-averaged gap $\langle \delta\rangle$, as a function of the ramp time $\alpha$ for $N=5-12$, with lighter to darker shades indicating increasing system size. Blue triangles denote the corresponding quench gaps, and black dots mark the peak gaps. Panels (b) and (d) show the dependence of the gap on the plateau duration $\kappa$, evaluated at the gap-maximizing $\alpha$. Dashed horizontal lines indicate the gap averaged over \(\kappa \in [10^{2},10^{5}]\). The insets in (a) and (c) show the system-size dependence of the peak gap (black circles), the $\kappa$-averaged gap (squares), and the quench gap (blue triangles). Fits to the exponential scaling of the inset data, reported in Table~\ref{tab:fits_k}, demonstrate improved scaling relative to the quench strategy for both models.}
    \label{fig:combine}
\end{figure*}

Fig.~\ref{fig:combine}(a) and (c) show the average spectral gap $\langle \delta \rangle$ at $\beta=5$ as a function of ramp time $\alpha$ for the SK and 3-spin models, respectively. The gap is obtained by exact diagonalization of the transition matrix at $h=1.5$ for system sizes $5\le N \leq 12$, followed by averaging over disorder realizations. The disordered-averaged gap is considered to probe the algorithm's generic behaviour without requiring instance-specific tuning. The phenomenology is similar to the Ising case, where a broad range of $\alpha$ improves the spectral gap relative to the quench protocol. The value of $\alpha$ that maximizes the spectral gap, indicated by the black dots, increases roughly linearly with system size, as examined in more detail in Appendix~\ref{appendix:scaling}. The inset shows the system-size dependence of the peak gap, with additional data for $ N=13$ and $14$, highlighting improved scaling relative to the quench case. 

Since these results correspond to the large-$\kappa$ limit, the dependence on the plateau duration $\kappa$, evaluated at the optimal $\alpha$, is shown separately in Fig.~\ref{fig:combine}(b) and (d). Although short-time oscillations are present, the gap remains close to its asymptotic value, suggesting that precise tuning of $\kappa$ is unnecessary. The scaling of the gap averaged over $\kappa \in [10^{2},10^{5}]$ is additionally compared in the insets of Fig.~\ref{fig:combine}(a) and (c). The large-$\kappa$ limit upper bounds the $\kappa$-averaged gap \cite{orfi2024barriers}, and the two are found to exhibit very similar scaling. The fitted exponential scaling exponents for the peak, $\kappa$-averaged and quench gaps, are listed in Table~\ref{tab:fits_k} for both spin-glass models. Additional analysis of the robustness and features of this scaling is presented in Appendix~\ref {appendix:scaling}, e.g. a simple schedule $\alpha \propto N$ produces a gap scaling that's close to the optimal schedule. The transverse field considered here, $h = 1.5$, does not correspond to the optimal quench performance \cite{orfi2024barriers}. Nevertheless, for both the SK and 3-spin models, the adiabatically dressed protocol exhibits improved scaling over the quench strategy for any value of $h$ \cite{orfi2024barriers}.

\begin{table}[t]
\centering
\renewcommand{\arraystretch}{1.15}
\setlength{\tabcolsep}{12pt}
\begin{tabular}{lcc}
\toprule
\multicolumn{3}{c}{Fitted Scaling Exponents} \\
\midrule
Protocol & SK & 3-spin \\
\midrule
Peak Gap                & \(0.146(2)\)  & \(0.182(3)\) \\
\(\kappa\)-Averaged Gap & \(0.124(1)\)  & \(0.181(1)\) \\
Quench Gap ($\alpha =0$)               & \(0.614(3)\) & \(0.453(3)\) \\
\bottomrule
\end{tabular}
\caption{Fitted exponential scaling exponents \(k\), defined by \(\langle \delta \rangle \propto 2^{-kN}\), for the peak, \(\kappa\)-averaged, and quench gaps of the SK and 3-spin models shown in Fig.~\ref{fig:combine}.}
\label{tab:fits_k}
\end{table}

\section{Conclusion}
In this work, we introduce an adiabatically dressed quantum proposal that improves on earlier quantum-enhanced MCMC methods while preserving their error resilience and near-term applicability. By replacing the quench with a time-symmetric ramp, the protocol suppresses the high-energy proposals generated by a sudden quench while retaining the nonlocality of the quantum update. The result is a broad intermediate regime between the quench and adiabatic limits in which the chain mixes substantially faster. 

In the Ising chain, bottleneck analysis enables the study of large system sizes and provides analytic access to the protocol’s performance. We demonstrate that, in contrast to the exponential decay of the gap observed in the quench limit, a ramped protocol yields an algebraic scaling of the gap with system size. These results show that a polynomial-time adiabatic dressing can circumvent the performance limitations revealed in previous work~\cite{orfi2024barriers}. 

We further extend our analysis to more challenging infinite-range models, namely the SK and 3-spin models, for which mixing is expected to remain exponentially slow. In both cases, the adiabatically dressed protocol substantially reduces the corresponding scaling exponent over the accessible range of system sizes. This improvement is robust to details of the ramp schedule and to the precise prescription used to choose the ramp time. It also substantially alleviates the need to fine-tune the strength of the quantum mixing term and evolution time, making the method significantly easier to implement experimentally. These features suggest that the protocol may offer practical advantages for large-scale optimization problems on near-term quantum hardware. 

\emph{Acknowledgements}
 A.O. acknowledges support from the Pierre Hohenberg Graduate Scholar Fellowship and computational resources provided by the Flatiron Institute. The Flatiron Institute is a division of the Simons Foundation. W.T.H. and D.S. were supported by AFOSR under Award No. FA9550-21-1-0236.

\clearpage
\bibliography{library}

\clearpage
\appendix
\setcounter{equation}{0}
\renewcommand{\theequation}{A\arabic{equation}}
\setcounter{figure}{0}
\renewcommand{\thefigure}{A\arabic{figure}}

\section{Ramp Functions}
\label{appendix:ramps}
\begin{figure}[h]
    \centering
    \begin{tikzpicture}
\begin{axis}[
    width=8cm,height=3.5cm,
    axis lines=left,
    xmin=0,xmax=6.2,
    ymin=0,ymax=1.25,
    xlabel={$t$},
    ylabel={$\gamma(t)$},
    xlabel style={
        at={(axis description cs:1,0)},
        anchor=west
    },
    xtick=\empty,
    ytick={0,1},
    yticklabels={$0$,$1$},
    clip=false,
]
\def\a{1.0}  
\def\k{4.0}  
\def\T{6.0} 
\addplot[thick,blue,domain=0:\a]
({x},{x/\a});
\addplot[thick,blue,domain=\a:\a+\k]
({x},{1});
\addplot[thick,blue,domain=\a+\k:\T]
({x},{(\T-x)/\a});
\addplot[dashed] coordinates {(\a,0) (\a,1)};
\addplot[dashed] coordinates {(\a+\k,0) (\a+\k,1)};
\draw[<->] (axis cs:0,-0.18) -- (axis cs:\a,-0.18)
    node[midway,below] {$\alpha$};
\draw[<->] (axis cs:\a,-0.18) -- (axis cs:\a+\k,-0.18)
    node[midway,below] {$\kappa$};
\draw[<->] (axis cs:\a+\k,-0.18) -- (axis cs:\T,-0.18)
    node[midway,below] {$\alpha$};
\end{axis}
\end{tikzpicture}
    \caption{The Hamiltonian parameter $\gamma(t)$ is linearly ramped from $0$ to $1$ over a duration $\alpha$ is followed by a plateau of length $\kappa$ and a time-reversed ramp back to $0$. Spectral gaps using this protocol in the large-$\kappa$ limit are shown in Fig.~\ref{fig:linearramp}.}
    \label{fig:linearrampdiagram}
\end{figure}
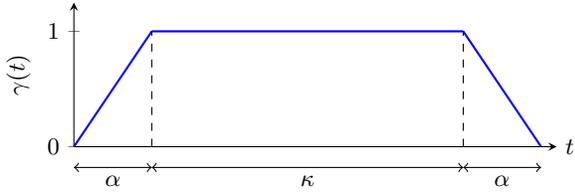

\begin{figure}[t]
    \centering   
    \begin{overpic}{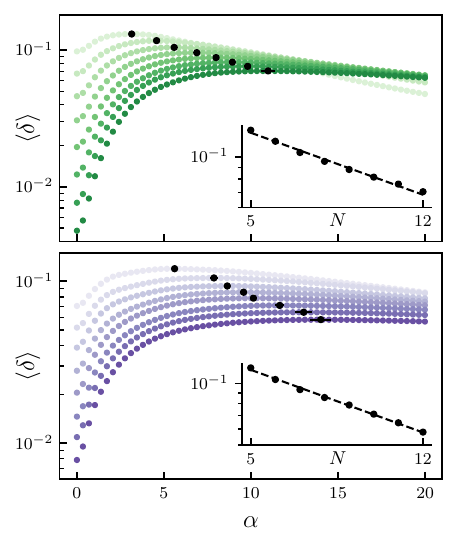}
    \put(74, 93){\fontsize{10}{12}\selectfont\textbf{(a)}}
    \put(74, 49.4){\fontsize{10}{12}\selectfont\textbf{(b)}}
\end{overpic}
    \vspace{-10pt}
    \caption{
    Spectral gap of the linear ramp protocol versus ramp time for the SK (a) and 3-spin (b) model at fixed transverse field $h=1.5$, showing similar behaviour to that in Fig.~\ref{fig:combine}. The inset shows the system-size dependence of the peak gap (black circles), from which the exponential scaling exponents $k_{peak}=0.129(3)$ for the SK model and $k_{peak}=0.146(3)$ for the 3-spin model are extracted.
    }  
    \label{fig:linearramp}
\end{figure}

To demonstrate the method's robustness across different ramp protocols, we also consider a linear ramp, shown in Fig.~\ref{fig:linearrampdiagram}. Although simpler to implement experimentally, this protocol may enhance diabatic transitions near the beginning and end of the ramp as compared to the doubly composed sine-squared ramp of Eq.~\ref{eq:ramp}. The spectral gap $\delta$ as a function of the ramp time, as well as the peak scaling, is shown in Fig.~\ref{fig:linearramp}. The spectral gap displays similar behavior to that presented in Fig.~\ref{fig:combine}, with only slight variations for small system sizes. In particular, the peak gap occurs at a system size dependent ramp time, and the adiabatic protocol still provides a clear enhancement over the quench limit, with scaling with system size showing slightly improved performance over the other ramp method.

\section{Ising Upper Bound}
\label{appendix:ising_UB}
As derived in Ref.~\cite{orfi2024barriers}, a tight upper bound on the spectral gap of the Ising chain is obtained by choosing the subset \(B\) in Eq.~\eqref{eq:lambda} to correspond to the first-excited state manifold. Specifically, let \(S_k\) denote the subset of configurations such that \(H_c(x)=E_k\) for all \(x\in S_k\), where \(E_k=-N+4k\) for the Ising chain. The resulting bottleneck upper bound then reads,
\begin{equation*}
    \delta \leq \frac{1}{N(N-1)}\sum_{x\in S_1,y\in S_0} Q(y|x) + \frac{e^{-4\beta}}{2-N(N-1)e^{-4\beta}}.
\end{equation*}
This bound can be simplified through a Jordan-Wigner transformation, allowing the Hamiltonian to be expressed in terms of momentum space fermionic operators, 
\begin{align*}
    H_p &= \sum_{K_p}\Big[\left(\gamma(t)h-\cos(k)\right)\left(a_k^\dagger a_k-a_{-k}a_{-k}^\dagger\right)\\
    &+i\sin(k)\left(a_{-k} a_{k}-a_k^\dagger a_{-k}^\dagger\right)\Big].
\end{align*}
The allowed momenta are
\begin{equation*}
K_p=\left\{\pm\frac{2\pi}{N}\left(l-\frac{p}{2}\right)\,\middle|\,
\begin{array}{l}
l=1,\ldots,N/2 \quad (p=0),\\
l=1,\ldots,N/2-1 \quad (p=1),
\end{array}
\right\}.
\end{equation*}
Since the momenta occur in pairs \((k,-k)\), the Hamiltonian can be written as a sum over positive momenta \(K_p^+\),
\begin{equation*}
    H_{0} = \sum_{k\in K_0^+} H_k, \qquad
    H_{1} = \sum_{k\in K_1^+} H_k + H_{k=0,\pi}.
\end{equation*}
Here the \(k=0\) and \(k=\pi\) modes in the \(p=1\) sector are treated separately,
\begin{equation*}
    H_{k=0,\pi}
    =2(\hat{n}_\pi-\hat{n}_0)
    +2\gamma(t)h(\hat{n}_0+\hat{n}_\pi-1).
\end{equation*}
Each mode \(H_k\) can be written in terms of $2\times 2$ matrix $\mathbf{H}_k$,
\begin{align*}
    H_k
    &= 2
    \begin{pmatrix}
        a_k^\dagger & a_{-k}
    \end{pmatrix}
    \mathbf{H}_k(t)
    \begin{pmatrix}
        a_k \\ a_{-k}^\dagger
    \end{pmatrix}, \\
    \mathbf{H}_k(t)
    &=2
    \begin{pmatrix}
        \gamma(t)h-\cos k & -i\sin k \\
        i\sin k & -\gamma(t)h+\cos k
    \end{pmatrix}.
\end{align*}
Since the momentum blocks are independent, the evolution factorizes as \(U=\prod_k U^k\). As in the quench case, the parity sectors decouple, yielding
\begin{align*}
\sum_{x\in S_1, y\in S_0} |\langle y|U|x\rangle|^2
&= \sum_{x\in S_1^{p=0}} |\langle gs_0|U|x\rangle|^2\\
&\quad + \sum_{x\in S_1^{p=1}} |\langle gs_1|U|x\rangle|^2 .
\end{align*}
The classical ground state is the Bogoliubov vacuum
\begin{equation*}
    \ket{gs_0}=\prod_{k\in K_0^+}\ket{0_k^c}, \quad \ket{0_k^c} = (\cos(k/2),-i\sin(k/2))
\end{equation*}
A classical excited state corresponds to single-particle excitation at some momentum \(k_i\),
\begin{equation*}
    \ket{x_{k_i}}=\ket{1_{k_i}^c}\prod_{k\neq k_i}\!\ket{0_k^c}, \quad \ket{1_{k}^c} = (\sin(k/2),i\cos(k/2))
\end{equation*}
The resulting sum becomes, 
\begin{align*}
    \sum_{x\in S_1,y\in S_0}|\langle y|U|x\rangle|^2&=
\sum_{k\in K_0^+}|\langle gs_0|U|x_k\rangle|^2\\
&+\sum_{k\in K_1^+}|\langle gs_1|U|x_k\rangle|^2
\end{align*}
where we have, 
\begin{equation*}
    \langle gs_p|U|x_{k_1}\rangle =
\langle 0_{k_1}^c|U^{k_1}|1_{k_1}^c\rangle
\prod_{k\neq k_1}\langle 0_k^c|U^k|0_k^c\rangle
\end{equation*}
Decompose \(U^k=U_3^kU_2^kU_1^k\) corresponding to the ramp up, evolution for time $\kappa$ with a fixed Hamiltonian, and then a ramp down. During $U_2^k$, \(H_h^k\) has eigenstates \(\ket{0_k},\ket{1_k}\) with energies  \(\pm \varepsilon_k=\pm \sqrt{(h-\cos k)^2+\sin^2 k}\). In the large-\(\kappa\) (time-averaged) limit the oscillatory interference terms vanish, giving,
\begin{equation*}
    |\bra{gs_p}U\ket{x_{k_i}}|^2 \simeq \left(
    |C_{k_i}|^2 + |D_{k_i}|^2
   \right)
   \prod_{k\neq k_i}
   \left(
    |A_k|^2 + |B_k|^2
   \right)
\end{equation*}
where 
\begin{align*}
A_k &= \bra{0_k^c}U_3^k\ketbra{0_k}U_1^k\ket{0_k^c}, \\
B_k &= \bra{0_k^c}U_3^k\ketbra{1_k}U_1^k\ket{0_k^c},\\
C_k &= \bra{1_k^c}U_3^k\ketbra{0_k}U_1^k\ket{0_k^c}, \\
D_k &= \bra{1_k^c}U_3^k\ketbra{1_k}U_1^k\ket{0_k^c}.
\end{align*}
Using the symmetry relation \(U_3^k=(U_1^{-k})^T\), all coefficients can be expressed in terms of a single transition probability $\eta_k = |\bra{0_k^c}U^k_1\ket{1_{k}}|^2$, 
\begin{align*}
    |A_k|^2 &=(1-\eta_k)^2, \\
    |B_k|^2 &= \eta_k^2,\\
    |C_k|^2 &= \eta_k(1-\eta_k),\\
    |D_k|^2 &= \eta_k(1-\eta_k).
\end{align*}
Defining $f_k=|A_k|^2+|B_k|^2=2\eta_k^2-2\eta_k+1$ the sum simplifies to
\begin{align*}
\sum_{k_i \in K_p^+} \left(|C_{k_i}|^2 + |D_{k_i}|^2\right)\prod_{k\neq k_i}\left(|A_k|^2 + |B_k|^2\right)\\
=
\Big(\prod_{k\in K_p^+}f_k\Big)
\sum_{k_i\in K_p^+}\!\left(\frac{1}{f_{k_i}}-1\right).
\end{align*}
This upper bound allows large systems to be probed numerically, as in Fig.~\ref{fig:ising_sin_scaling}, as it only requires $\eta_k$ to be estimated numerically instead of the entire time-dependent evolution. In the large $N$ limit, this bound can be expressed as 
\begin{align*}
    \delta &\leq \tfrac{1}{\pi(N-1)}\exp\left(-\tfrac{N}{2\pi}\int_0^\pi\log\left(\tfrac{1}{f_k}\right) dk\right)\int_0^\pi\left(\tfrac{1}{f_k}-1\right) dk\\
    &\quad + \frac{e^{-4\beta}}{2-N(N-1)e^{-4\beta}}
    \label{eq:delta_isingf}
\end{align*}
For an arbitrary $U_1^k$, all dependence of the gap on the unitary is encoded in $f_k$. For small transition probabilities, $\log(1/f_k)\approx 2\eta_k$, so removing the exponential scaling requires $\int_0^\pi\eta_k dk = O(1/N)$. Since this also implies, $\int_0^\pi \left(\tfrac{1}{f_k}-1\right) dk = O(1/N)$, the gap cannot scale better than $O(1/N^2)$ for any choice of unitary.

\section{Ising Linear Ramp}
\label{appendix:ising_linear}
The transition probability $\eta_k$ for the Ising chain with a linear ramp can be found exactly as the dynamics factorize into a set of independent Landau Zener problems \cite{dziarmaga2005dynamics}. Specifically, the initial state $\ket{0}_k^c = (\cos(k/2),-i\sin(k/2))$ is evolved for a time $\alpha$ under the time-dependent Hamiltonian, 
\begin{equation*}
    \mathbf{H}_k(t) = 2\begin{bmatrix}
    -\cos(k)+\frac{h}{\alpha} t & -i\sin(k)\\
    i\sin(k) &cos(k)-\frac{h}{\alpha} t
\end{bmatrix} = H_0 +2\frac{h}{\alpha}t\sigma_z.
\end{equation*}
Projecting the evolved state onto the instantaneous excited eigenstate,
\begin{equation*}
    \ket{1}_{k} = (\cos(\theta_f/2),i\sin(\theta_f/2)) 
\end{equation*}
with $\theta_f = \tan^{-1}(\sin(k)/(h-\cos(k)))$, gives the desired transition probability $\eta_k$. 

Let $g= 2\sin(k)$, $\delta=2h/\alpha$ and $\Tilde{t} = t-\cos(k)\alpha/h$ which gives, 
\begin{equation*}
    \mathbf{H}_k(t) = g\sigma_y + \delta \Tilde{t}\sigma_z, 
\end{equation*}
the typical Landau-Zener form. The time-evolved state may be written as $U_1^k\ket{0}_k^c = (C_1,C_2)$, where the coefficients can be expressed in terms of parabolic cylinder functions \cite{vitanov1996landau}. Defining $z_i = -2e^{-i\pi/4}\cos(k)\sqrt{\alpha/h}$, $z_f = 2e^{-i\pi/4}\sqrt{\alpha/h}(h-\cos(k))$ and $\nu = i\sin^2(k)\alpha/h$ one finds,
\begin{align*}
    C_1(t_f) 
    &= -\sqrt{\frac{\alpha}{h}}\frac{i\sin^2(k)\cos(k/2)}{\sqrt{2\pi}}\Gamma(-\nu)\\
    &\quad \times \Bigg[\sqrt{\frac{\alpha}{h}}\mathcal{D}_\nu^{1} 
    + \frac{2\alpha}{h}e^{-i\pi/4}\sin^2(k/2)\mathcal{D}_\nu^{2}\Bigg]
\end{align*}
\begin{align*}
   C_2(t_f)  &= \frac{i\sin(k)\cos(k/2)}{\sqrt{2\pi}}\Gamma(-\nu)\\
   &\quad \times \Bigg[e^{i\pi/4}\sqrt{\frac{\alpha}{h}}\mathcal{D}_\nu^{3} +\frac{2\alpha}{h}\sin^2(k/2)\mathcal{D}_\nu^{4}\Bigg]
\end{align*}
where
\begin{align*}
    \mathcal{D}_\nu^{1} &=D_\nu(-z_i)D_{\nu-1}(z_f)+D_\nu(z_i)D_{\nu-1}(-z_f)\\
    \mathcal{D}_\nu^{2} &= D_{\nu-1}(-z_i)D_{\nu-1}(z_f)-D_{\nu-1}(z_i)D_{\nu-1}(-z_f)\\
    \mathcal{D}_\nu^{3} &= D_\nu(-z_i)D_{\nu}(z_f)-D_\nu(z_i)D_{\nu}(-z_f)\\
    \mathcal{D}_\nu^{4}&= D_{\nu-1}(-z_i)D_{\nu}(z_f)+D_{\nu-1}(z_i)D_{\nu}(-z_f)
\end{align*}
The transition probability can therefore be written as,
\begin{equation*}
    \eta_k = \left|\cos\left(\frac{\theta_f}{2}\right)C_1(t_f) -i \sin\left(\frac{\theta_f}{2}\right)C_2(t_f)\right|^2.
\end{equation*}

\subsection*{Large Transverse Field Approximation} 
The transition probability can be recast using the connection formula for the parabolic cylinder function \cite{NIST:DLMF},
\begin{equation*}
D_\nu(z)=e^{i\pi\nu}\,D_\nu(-z)+\frac{\sqrt{2\pi}}{\Gamma(-\nu)}\,e^{i\pi(\nu+1)/2}\,D_{-\nu-1}(-i z),
\end{equation*}
allowing the removal of the explicit gamma-function factors. Letting $l = \frac{\alpha}{h}$ to simplify notation, this gives
\begin{equation*}
    \eta_k = e^{-\pi l \sin^2(k)} \left|A\right|^2,
\end{equation*}
where 
\begin{align*}
    A&=\cos\!\left(\frac{\theta_f}{2}\right)\left[\cos\!\left(\frac{k}{2}\right)J+\sqrt{l}\,e^{-i\pi/4}\sin(k)\sin\!\left(\frac{k}{2}\right)K\right]\\
    &\quad+\sin\!\left(\frac{\theta_f}{2}\right)\left[\sqrt{l}\,e^{i\pi/4}\cos\!\left(\frac{k}{2}\right)\sin(k)L+\sin\!\left(\frac{k}{2}\right)M\right],
\end{align*}
and the coefficients are defined as,
\begin{align*}
    J &=-l\sin^{2}(k)\,D_{-\nu-1}(iz_i)\,D_{\nu-1}(z_f)- D_{\nu}(z_i)\,D_{-\nu}(iz_f),\\
    K &=D_{\nu-1}(z_i)\,D_{-\nu}(iz_f)- D_{-\nu}(iz_i)\,D_{\nu-1}(z_f),\\
    L &=D_{-\nu-1}(iz_i)\,D_{\nu}(z_f) - D_{\nu}(z_i)\,D_{-\nu-1}(iz_f),\\
    M &=l\sin^{2}(k)\,D_{\nu-1}(z_i)\,D_{-\nu-1}(iz_f)+ D_{-\nu}(iz_i)\,D_{\nu}(z_f).
\end{align*}
To probe the scaling behaviour observed in Fig.~\ref{fig:ising_sin_scaling}, we approximate the parabolic cylinder functions in the large-$h$ limit. The explicit dependence on $h$ is examined in more detail in the following section. For sufficiently large $h$, the magnitude $|z_f|$ is large for all $k$, justifying the use of the leading-order large-$|z|$ asymptotic approximation of $D_\nu(z)$ \cite{NIST:DLMF},
\begin{equation*}
    D_{\nu}(z) \simeq e^{-z^2/4}z^{\nu}.
\end{equation*}
Removing sub-leading terms in $h$, the transition probability takes the simplified form,
\begin{align*}
    \eta_k &\simeq \cos^2(k/2)e^{-\frac{\pi}{2}l\sin^2(k)}\\
    &\quad \times\left|D_\nu(z_i) - 2\sqrt{l}\sin^2(k/2)e^{-i\pi/4}D_{\nu-1}(z_i)\right|^2
\end{align*}
Although this expression is significantly simpler than the exact result, it still involves special functions, which complicates a direct analysis of the algorithm’s scaling behaviour. As observed numerically in Fig.~\ref{fig:ising_sin_scaling}, the scaling becomes increasingly consistent in the large-$\alpha$ ($\mathcal{O}(N)$) regime, and we therefore focus on this limit in what follows. Unfortunately, the magnitude $|z_i|$ varies with $k$, making neither the large-$|z|$ nor small-$|z|$ approximations of $D_\nu(z_i)$ valid for all $k$. However, when $\alpha$ is larger, the support of $\eta_k$ is concentrated on smaller $k$, as the evolution becomes more adiabatic, high-energy excitations are less likely. This observation justifies considering the large-$|z|$ approximation of the parabolic cylindrical functions valid for small $k$ and neglecting contributions from larger $k$, thereby capturing the dominant scaling behaviour. This expansion, and approximating for small $k$ gives, 
\begin{equation*}
    \eta_k \simeq e^{-2\pi l\sin^2(k)}\cos^2(k/2)
\end{equation*}
and therefore,
\begin{equation*}
    f_k \simeq 1 -2e^{-2\pi lk^2} + 2e^{-4\pi lk^2} + k^2\left(\frac{1}{2}e^{-2\pi lk^2}-e^{-4\pi lk^2}\right).
\end{equation*}
As stated in Appendix~\ref{appendix:ising_UB}, the exponential term of the bottleneck bound depends on the following integral. Keeping the leading small-$k$ terms gives (using that in this regime $l>1)$,
\begin{align*}
    -\int_0^\pi &\log(f) dk \simeq-\int_0^\pi\log( 1 -2e^{-2\pi lk^2} + 2e^{-4\pi lk^2}) dk,\\
    &= -\frac{1}{\sqrt{2\pi l}}\int_0^{\pi \sqrt{2\pi l}}\log( 1 -2e^{-x^2} + 2e^{-2x^2}) dx,\\
    &\geq \frac{1}{\sqrt{2\pi l}}\int_0^{\infty} (2e^{-x^2} - 2e^{-2x^2}) dx,\\
    &= c\sqrt{\frac{h}{\alpha}}.
\end{align*}

Therefore, if $\alpha$ is scaled as $\mathcal{O}(N^2)$, the exponential term is removed. As compared to the exact upper bound, this approximation underestimates the gap at large $\alpha$, a smaller polynomial scaling of $\alpha$ may still result in polynomial scaling of the spectral gap. The remaining term in the bottleneck bound gives,
\begin{align*}
    \int_0^\pi\left(\frac{1}{f}-1\right)dk &\simeq\int_0^\pi\frac{2e^{-2\pi l k^2}-2e^{-4\pi l k^2}}{1-2e^{-2\pi l k^2}+2e^{-4\pi l k^2}}dk,\\
    &= \frac{1}{\sqrt{2\pi l}}\int_0^{\pi\sqrt{2\pi l}}\frac{2e^{-x^2}-2e^{-2x^2}}{1-2e^{-x^2}+2e^{-2x^2}}dx,\\
    &< \frac{1}{\sqrt{2\pi l}}\int_0^{\infty}\frac{2e^{-x^2}-2e^{-2x^2}}{1-2e^{-x^2}+2e^{-2x^2}}dx,\\
    &= C\sqrt{\frac{h}{\alpha}}.
\end{align*}
Putting this together with the prefactor gives the expected $O(N^{-2})$ scaling of the spectral gap.

\subsection*{Small Ramp Time Approximation} 
To probe the effect of a time-dependent protocol as a function of the transverse field, we focus on the small $\alpha$ regime to determine when introducing a ramp improves performance relative to the quench proposal. In this limit, the time-dependent evolution can be approximated with a Magnus expansion, 
\begin{align*}
    U_1 &= \mathcal{T}\exp\left(-i\int_0^\alpha H(t) dt\right)\simeq  \exp\left(\sum_i\Omega_i\right),
\end{align*}
where the operators $\Omega_i$ are given by time-ordered nested commutators. Recall, for the Ising case with a linear ramp, we have the following time-dependent Hamiltonian, 
\begin{align*}
    H_k(t)
    &=  2\left(\sin(k)\sigma_y -\cos(k)\sigma_z \right) +\frac{t}{\alpha} 2h\sigma_z.
\end{align*}
Defining, $\sum_i\Omega_i
=i\big(a_x\sigma_x+a_y\sigma_y+a_z\sigma_z\big)$, the expansion truncated at second order gives,
\begin{align*}
a&=\Big(\tfrac{2}{3}h\alpha^2\sin k,\,-2\alpha\sin k,\,\alpha(2\cos k-h)\Big).
\end{align*}
The corresponding transition amplitude up to order $\alpha^2$ is,
\begin{align*}
\eta_k &=\Bigl|\cos|a|\,\cos\left(\tfrac{\theta_k+k}{2}\right)\\
    +\tfrac{\sin\!|a|}{|a|}&\Bigl(a_x\sin\!\Bigl(\tfrac{\theta_k+k}{2}\Bigr)
    +i a_y\sin\!\Bigl(\tfrac{\theta_k-k}{2}\Bigr)
    +i a_z\cos\!\Bigl(\tfrac{\theta_k-k}{2}\Bigr)\Bigr|^2, \\
    \simeq  &\cos^2\left(\frac{\theta_k+k}{2}\right)- \frac{\alpha^2h^2\sin^2(k)}{3\varepsilon_k}.
\end{align*}
Where here we used $\theta_k = \cos^{-1}\left(\frac{h-\cos(k)}{\varepsilon_k}\right)$ with $\varepsilon_k = \sqrt{(h-\cos(k))^2+\sin^2(k)}$. For the quench case, $\eta_k^0 := \cos^2\left(\frac{\theta_k+k}{2}\right)$ which gives $f_k^0  = 1-\frac{h^2\sin^2(k)}{2\varepsilon_k^2} $. To determine the change from the quench case, define $\eta_k = \eta_k^0 - c_k$, which gives,
\begin{align*}
    f_k &= f_k^0+2c_k^2-4\eta_k^0c_k + 2c_k,\\
    &= f_k^0 + g_k.
\end{align*}
As $\alpha$ is small, we can expand the terms needed for the upper bound,
\begin{align*}
    &\left(\prod_{k} f_{k}\right) \sum_{k_i}\left(\frac{1}{f_{k_i}}-1\right)\\
    &= \left(\prod_{k} (f^0_{k}+g_k)\right) \sum_{k_i}\left(\frac{1}{(f^0_{k_i}+g_{k_i})}-1\right),\\
    &\simeq \prod_{k} f^0_{k} \left(1+ \sum_k\frac{g_{k}}{f_{k}^0}\right)\left[ \sum_{k_i}\left(\frac{1}{f^0_{k_i}}-1\right) - \sum_{k_i}\frac{g_{k_i}}{(f_{k_i}^0)^2}\right],\\
    &= \left(\prod_{k} f_{k}^0\right) \sum_{k_i}\left(\frac{1}{f_{k_i}^0}-1\right) +\left(\prod_{k} f_{k}^0\right)\left[\sum_k g_k \frac{Sf_k^0-1}{(f_k^0)^2}\right],
\end{align*}
where $ S = \sum_{k}\left(\frac{1}{f_{k}^0}-1\right)$. We focus on the sign of the second term to determine whether the result represents an increase or a decrease relative to the quench value. Here $S$ is always positive, and as $\frac{1}{2}\leq f_0\leq 1$ the product $\prod_{k} f_{k}^0$ is also positive. Additionally, if we consider the large $N$ limit, 
\begin{align*}
    S &= \frac{N}{2\pi}\int_0^\pi dk\left(\left(1-\tfrac{h^2\sin^2(k)}{2\varepsilon_k^2}\right)^{-1} -1\right),\\
    &= \begin{cases}
        \frac{N}{2}\left(\sqrt{\frac{2}{2-h^2}}-1\right), & \text{if } 0\leq h \leq 1\\
        \frac{N}{2}\left(\sqrt{2}-1\right), & \text{if }  h > 1
    \end{cases}
\end{align*}
If we restrict to considering the situation where $h\leq1$, we have,
\begin{align*}
     \frac{1}{(f_k^0)^2}&g_k(Sf_k^0-1)\\
     &=\frac{1}{(f_k^0)^2} \tfrac{2\alpha^2h^2\sin^2(k)}{3\varepsilon_k^2}\left(1-h\cos(k)\right)(Sf_k^0-1),\\
     \leq \tfrac{1}{(f_k^0)^2}& \tfrac{2\alpha^2h^2\sin^2(k)}{3\varepsilon_k^2}\left(1-h\cos(k)\right)\left(\frac{N}{2}\left(\sqrt{\tfrac{2}{2-h^2}}-1\right) -1\right).
\end{align*}
Therefore, the correction term is negative for all $k$ if $h\leq \sqrt{8(N+1)}/(N+2)\simeq \sqrt{8/N}$. This corresponds to the regime where $h$ is smaller than the optimal quench value of $h$, where the quantum eigenstates remain close to the classical basis states. 

To show this precisely, we can derive the optimal quench $h$ from the bottleneck bound. The exponential term can be bounded using Jensen's inequality, 
\begin{align*}
    \int_0^\pi\log\left(\frac{1}{f_k^0}\right) dk &= - \int_0^\pi\log\left(1-\frac{h^2\sin^2(k)}{2\varepsilon^2_k}\right)dk,\\
    &\geq -\pi\log\left(1-\frac{1}{\pi}\int_0^\pi\frac{h^2\sin^2(k)}{2\varepsilon^2_k}dk\right),\\
    &= \begin{cases}
        \pi\log\left(\frac{4}{4-h^2}\right), & \text{if } |h|\leq 1\\
        \pi\log\left(\frac{4}{3}\right), & \text{if }  |h| \geq 1
    \end{cases}
\end{align*}
For the regime we are considering $|h|\leq 1$, the quench bottleneck bound is,
\begin{align*}
    \delta &\leq\frac{\sqrt{\tfrac{2}{2-h^2}}-1}{N-1}\left(1-\tfrac{h^2}{4}\right)^{N/2} + \frac{e^{-4\beta}}{2-N(N-1)e^{-4\beta}}.
\end{align*}
This expression is maximized to leading order at $h\simeq \sqrt{8/N}$. Consequently, the time-dependent proposal offers an advantage only for transverse fields exceeding the optimal quench value, consistent with the intuition based on eigenstate delocalization.

\section{General Landau-Zener Approximation} 
\label{appendix:ising_LZ}
Without considering the details of the ramp protocol, some insight into the Markov chain's performance on the Ising model can be obtained from the general form of the Landau-Zener transition probability. For each momentum $k$ mode, 
\begin{equation*}
    \eta_k \simeq  e^{-\frac{2\pi}{h}\alpha\Delta_k^2}
\end{equation*}
where $\Delta_k$ denotes the minimum gap associated with that mode \cite{dziarmaga2005dynamics}. When $h>1$, the evolution crosses the phase transition, and the low-momentum modes dominate the excitation probability. As the Ising transition has a dynamical critical exponent $z=1$, the minimal gap scales linearly with momentum, $\Delta_k \sim k$. This gives the approximation, 
\begin{equation*}
    \eta_k \simeq  e^{-\frac{2\pi}{h}\alpha k^2},
\end{equation*}
in agreement with the approximation derived in Appendix~\ref{appendix:ising_linear}. Recall that polynomial scaling requires, 
\begin{equation*}
    \int_0^\pi\log\left(f_k\right) dk \sim \frac{1}{N}.
\end{equation*}
Using the approximation $\log(f_k)\sim -2\eta_k$ for small $\eta_k$, 
\begin{align*}
    \int_0^\pi\log\left(f_k\right) dk &\simeq -2\int_0^\pi e^{-\frac{2\pi}{h}\alpha k^2} dk, \\
    &\simeq  -\sqrt{\frac{h}{2\alpha}},
\end{align*}
it follows that achieving polynomial scaling requires $\alpha$ to scale as $\mathcal{O}(N^2)$, consistent with the result derived above.

\begin{figure}[h]
    \centering
    \includegraphics[width=\linewidth, trim=0.2cm 0.32cm 0 0.1cm, clip]{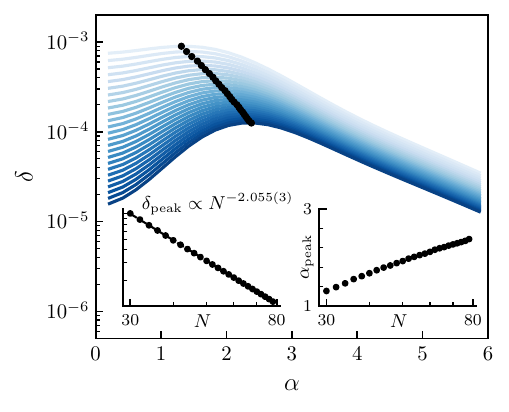}
    \caption{At $h=2/3$, the bottleneck bound on the spectral gap $\delta$ for the Ising chain is shown as a function of the ramp time. Insets show the polynomial scaling of the peak gap (left) and the logarithmic scaling of the ramp time $\alpha_{\rm peak}$ at which the peak gap is attained (right).}
    \label{fig:ising_2-3}
\end{figure}

If the phase transition is not crossed, each $k$ mode remains gapped throughout the evolution, and we can assume $\Delta_k$ has weak $k$ dependence, 
\begin{align*}
    \int_0^\pi\log\left(f_k\right) dk &\simeq -2\int_0^\pi e^{-\frac{2\pi}{h}\alpha\Delta_k^2} dk, \\
    &\simeq  -2\pi e^{-\frac{2\pi}{h}\alpha\Delta_k^2}.
\end{align*}
Under these assumptions, polynomial scaling can be achieved if $\alpha$ scales as $\mathcal{O}(\log(N))$. This behaviour is confirmed numerically in Fig.~\ref{fig:ising_2-3}, which shows the spectral gap at $h=2/3$ for the smooth ramp of Eq.~\ref{eq:ramp}, analogous to Fig.~\ref{fig:ising_sin_scaling}. The peak spectral gap maintains polynomial scaling with system size, while the value of $\alpha$ at which the peak occurs grows only logarithmically. 

As discussed in the previous section, the time-dependent proposal can outperform the quench strategy when the transverse field exceeds the optimal quench value. In this model, the optimal quench transverse field is below the critical point as there exists a local update strategy (spin flips) that produces a polynomial scaling Markov chain \cite{orfi2024barriers}. As a result, the improved scaling remains accessible for $h<1$, as shown in Fig.~\ref{fig:ising_2-3}. The time-dependent protocol is able to achieve this scaling without crossing the phase transition due to the locality of the optimal proposals. It remains unclear whether the same is true in more complex systems, where efficient mixing may require non-local updates. 

When the phase transition is crossed, as in Fig.~\ref{fig:ising_sin_scaling}, there is a broad range of $\alpha$ over which the spectral gap remains near its maximum. By contrast, for smaller $h$, the system remains gapped throughout the evolution so the onset of the overly adiabatic behaviour begins at smaller $\alpha$, leading to a narrower region of optimal performance. There's thus a benefit in crossing the critical point as a broad optimal range of $\alpha$ makes the system more robust to changes in the ramp time.

\section{Spin Glass Scaling Properties}
\label{appendix:scaling}
\begin{figure}[h]
    \centering
    \includegraphics[width=\linewidth, trim=0.2cm 0.32cm 0 0.1cm, clip]{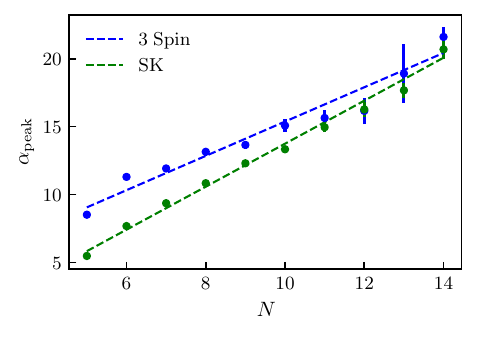}
    \caption{System-size scaling of $\alpha_{\rm peak}$, the value of $\alpha$ that maximizes the instance-average spectral gap in Figs.~\ref{fig:combine}(a) and \ref{fig:combine}(c), showing linear behaviour for both the SK and 3-spin models.}
    \label{fig:alpha_scaling}
\end{figure}

Fig.~\ref{fig:combine} shows the system-size scaling of the instance-averaged spectral gap evaluated at $\alpha_{\rm peak}$, where $\alpha_{\rm peak}$ is the value of $\alpha$ that maximizes the average gap. As shown in Fig.~\ref{fig:alpha_scaling}, $\alpha_{\rm peak}$ grows approximately linearly with system size for both the SK and 3-spin models. For the large system sizes considered, the spectral gap plateaus, leading to larger uncertainty in the position of $\alpha_{\rm peak}$.

\begin{figure}[h]
    \centering
    \includegraphics[width=\linewidth, trim=0.2cm 0.32cm 0 0.1cm, clip]{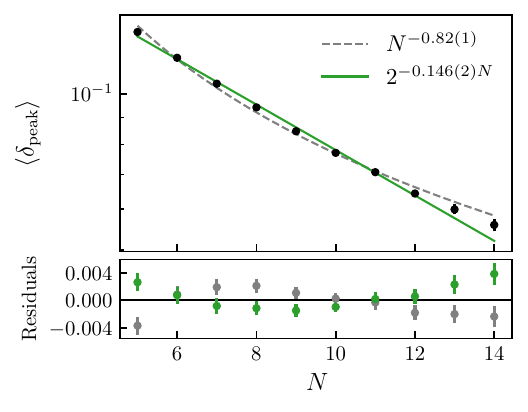}
    \caption{System-size dependence of the SK peak gap with power-law (grey dashed) and exponential (green solid). The lower panel shows the corresponding residuals, revealing systematic deviations at the largest $N$ accessible. The exponential fit is modestly favoured by the reduced chi-square, $\chi^2_\nu=2.42$ compared with $\chi^2_\nu = 3.02$ for the power-law fit.}
    \label{fig:SK_scaling_details}
\end{figure}

For the limited range of $N$ accessible here, the SK peak gap does not exactly follow a clean exponential trend at the largest system sizes. Fig.~\ref{fig:SK_scaling_details} compares power-law and exponential fits to the peak gap scaling. Both functional forms show deviations at the largest $N$ considered. The power-law fit yields a reduced chi-square $\chi^2_\nu = 3.02$, while the exponential fit gives $\chi^2_\nu=2.42$, favouring the exponential fit reported in the main text.

\begin{figure}[t]
    \centering
    \includegraphics[width=\linewidth, trim=0.1cm 0.32cm 0 -0.2cm, clip]{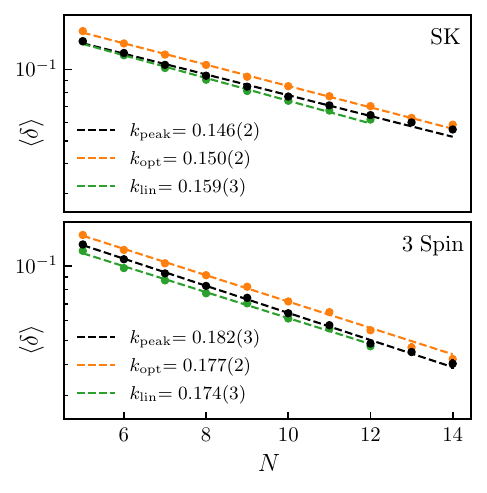}
    \caption{Comparison of the scaling exponents $k$ extracted under three prescriptions for the ramp time ramp time $\alpha$. The main-text exponent $k_{\rm peak}$ is obtained from the peak of the disorder-averaged gap (black). The instance optimized benchmark $k_{\rm opt}$ is obtained by fitting the average of the instance-wise optimal gaps (orange). Finally, $k_{\rm lin}$ sets $\alpha=N$, and is shown only for $N\leq 12$, where a large $\alpha$ scan is available (green). The close agreement among the three exponents indicates that the scaling is largely insensitive to the specific method used to choose $\alpha$.}
    \label{fig:scaling_compare}
\end{figure}

As noted in the main text, to capture the algorithm's generic performance the focus is on the peak of the disorder instance averaged gap, yielding the scaling exponent $k_{\rm peak}$. Fig.~\ref{fig:scaling_compare} additionally compares the scaling exponent $k_{\rm opt}$ obtained by averaging the instance-wise optimal gap. This benchmark is not practically feasible, as it presumes prior knowledge of the optimal ramp time $\alpha$ for each Hamiltonian instance. Finally, we investigate the other extreme, where we avoid explicit maximization and instead adopt a simple strategy, $\alpha = N$, which requires no prior knowledge of the optimal $\alpha$. This linear choice of $\alpha = N$ is only shown up to $N=12$, corresponding to the data in Figs.~\ref{fig:combine}. Across both models, the three extracted scaling exponents are similar and smaller than the best quench exponents reported in Ref.~\cite{orfi2024barriers}, indicating that the method is robust to the specific prescription for choosing $\alpha$. As seen in Figs.~\ref{fig:combine}, once $\alpha$ exceeds the initial gap enhancement, the gap depends only weakly on $\alpha$. This performance insensitivity to protocol details suggests that the method is well-suited to large-scale experiments.

For the SK model, the departures from exponential scaling at the largest $N$ seen in Fig.~\ref{fig:SK_scaling_details} are less pronounced for the instance optimized gap values. Specifically, the peak gap and optimal gap values converge as $N$ increases, consistent with reduced instance-to-instance variability at larger sizes. This may account for some of the deviations observed in Fig.~\ref{fig:SK_scaling_details}.

\end{document}